\documentclass[
 ,final            
  ]
  {aipproc}

\layoutstyle{6x9}
\newcommand{\aap}{A\&A}
\newcommand{\apj}{ApJ}
\newcommand{\mnras}{MNRAS}

\newcommand{\aj}{AJ}
\newcommand{\nat}{Nature}
\newcommand{\beq}{\begin{equation}}
\newcommand{\eeq}{\end{equation}}
\newcommand{\beqn}{\begin{eqnarray}}
\newcommand{\eeqn}{\end{eqnarray}}

\newcommand{\pd}{\partial}

\begin{document}

\title{Particle acceleration efficiencies in astrophysical shear flows}

\author{Frank M. Rieger \,and\, Peter Duffy}{
  address={Department of Mathematical Physics, University College Dublin, 
           Belfield, Dublin 4, Ireland}
}

\begin{abstract}
The acceleration of energetic particles in astrophysical shear flows is analyzed.
We show that in the presence of a non-relativistic gradual velocity shear, power 
law particle momentum distributions $f(p) \propto p^{-(3+\alpha)}$ may be 
generated, assuming a momentum-dependent scattering time $\tau \propto p^{\alpha}$, 
with $\alpha > 0$. We consider possible acceleration sites in astrophysical jets 
and study the conditions for efficient acceleration. It is shown, for example, 
that in the presence of a gradual shear flow and a gyro-dependent particle mean 
free path, synchrotron radiation losses no longer stop the acceleration once it 
has started to work efficiently. This suggests that shear acceleration may 
naturally account for a second, non-thermal population of energetic particles 
in addition to a shock-accelerated one. The possible relevance of shear acceleration
is briefly discussed with reference to the relativistic jet in the quasar 3C~273.
\end{abstract}

\maketitle


\section{Introduction}
Collimated relativistic outflows have been detected in a variety of sources 
including AGNs, microquasars and $\gamma$-ray bursts (e.g., \cite{mir99,zen97}). 
Today there is strong evidence that many (if not all) of these jets are 
characterized by a significant shear in their velocity field (cf. \cite{rie04} 
for a review). It appears thus very likely that energetic particles can be 
accelerated efficiently by scattering off magnetic inhomogeneities carried 
within such shear flows. 
Possible implications of such an acceleration process have so far been studied 
by several authors: in the pioneer works on non-relativistic gradual shear flows, 
Berezhko \& Krymskii~(1981), for example, showed that under suitable conditions 
power law particle momentum distributions may be formed (cf. also \cite{ber81}), 
whereas Earl, Jokpii \& Morfill~(1988) derived the corresponding diffusive 
particle transport equation including shear and inertial effects, assuming the 
diffusion approximation to be valid. The work by Earl et al. was generalized to 
the relativistic regime by Webb (cf. \cite{web85,web89,web94}) using a 
mixed-frame approach. Based on those results, Rieger \& Mannheim~(2002) have 
recently studied the acceleration of particles in rotating and shearing flows 
with application to the relativistic jets in AGN. Particle acceleration in 
non-gradual relativistic shear flows (e.g., in the presence of a relativistic 
velocity jump), on the other hand, has been studied by Ostrowski (\cite{ost90,
ost98,ost00}) using Monte Carlo simulations, showing that very flat momentum 
spectra may be possible.\\ 
In the present contribution we will particularly focus on the conditions required 
for efficient shear acceleration in order to assess whether shear-accelerated 
particles may indeed significantly contribute to the emission at a certain energy 
band.

\section{Basic principles of shear acceleration}
\begin{figure}
    \includegraphics[height=.18\textheight]{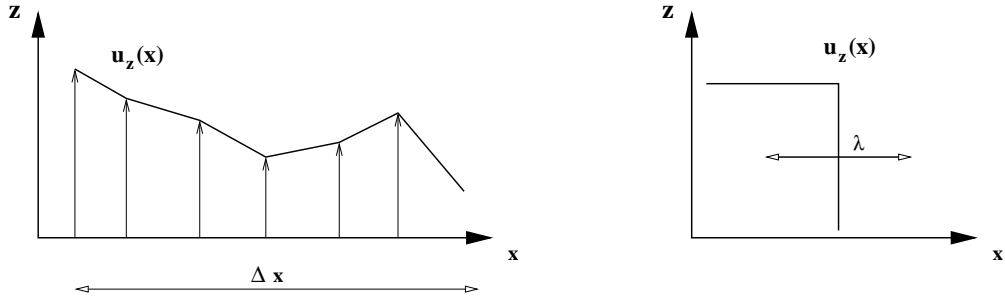}
    \caption{Sketch of a two-dimensional, non-relativistic shear velocity profile
     with the flow directed along the z-axis ({\it{left}}). Qualitatively, a particle 
     may be considered as experiencing a gradual shear profile if its mean free path 
     $\lambda$ is much smaller than the width $\Delta x$ of the shear layer and if 
     the flow velocity changes smoothly within that layer. In the case where the 
     particle becomes so energetic that its mean free path $\lambda = c\,\tau$ 
     becomes larger than $\Delta x$, it essentially experiences a non-gradual
     velocity shear, i.e. a discontinuous jump in the flow velocity ({\it{right}}).}
      \label{fig1}
  \end{figure}

Shear acceleration is based on the idea that energetic particles may gain energy 
by scattering off systematically moving small-scale magnetic field irregularities. 
These irregularities are thought to be embedded in a collisionless shear flow such 
that their velocities correspond to the local flow velocity, i.e. second-order 
Fermi effects are neglected. The scattering process is assumed to occur in such
a way that the particles are randomized in direction and their energies conserved 
in the local comoving fluid frame. In the presence of a velocity shear, the momentum 
of a particle travelling across the shear changes (with respect to the local fluid 
frame), so that a net increase may occur (e.g., \cite{jok90}). For illustration 
consider a non-relativistic continuous shear flow with velocity profile given by 
(cf. {\bf{Fig.~\ref{fig1}}})
\beq
   \vec{u} = u_z(x)\,\vec{e}_z\,.
\eeq  
Let $\vec{v}=(v_x,v_y,v_z)$ be the velocity vector, $m$ the relativistic mass and 
$\vec{p}_1$ the initial momentum (relative to local flow frame) of the particle. 
Within one scattering time $\tau$ (initially assumed to be independent of momentum) 
the particle travels a distance $\delta x = v_x\,\tau$ across the shear, so that
in the presence of a gradual shear the flow velocity will have changed by $\delta 
\vec{u} = \delta u\,\vec{e_z}$, with $\delta u = (\partial u_z/\partial x)\,\delta 
x$. Hence the particle's momentum relative to the flow becomes $\vec{p}_2 = 
\vec{p}_1 + m\,\delta \vec{u}$, i.e.
\beq\label{trafo}
    p_2^2 = p_1^2 + 2\,m\,\delta u\,\,p_{1,z} + m^2\,(\delta u)^2\;.
\eeq As the next scattering event preserves the magnitude of the particle momentum 
(in the comoving frame) the particle magnitude will have this value in the local 
flow frame and hence a net increase in momentum may occur with time. Using spherical 
coordinates and averaging over solid angle (assuming an almost isotropic particle 
distribution), the average rate of momentum change and the average rate of momentum 
dispersion can be written as (cf. \cite{jok90})
\beqn
  \left<\frac{\Delta p}{\Delta t}\right> & \equiv & \frac{2\,<(p_2 - p_1)>}{\tau}
                    =  \frac{4}{15}\,p \left(\frac{\pd u_z}{\pd x}\right)^2 \tau 
                     \label{p_dot}\\
  \left<\frac{\Delta p^2}{\Delta t}\right> & \equiv & \frac{2\,<(p_2 - p_1)^2>}{\tau} 
                    = \frac{2}{15}\,p^2 \left(\frac{\pd u_z}{\pd x}\right)^2 \tau \;,
\eeqn i.e. both depending on the square of the flow velocity gradient. Note that 
it can be shown that a momentum-dependent scattering time obeying a power law of the 
form $\tau \propto p^{\alpha}$ may be accommodated by replacing $4 \rightarrow (4 +
\alpha)$ in Eq.~(\ref{p_dot}). 
Using these results we may write down a simple Fokker-Planck transport equation for 
the phase space particle distribution function $f(p)$, assuming a mono-energetic 
injection of particles with momentum $p_0$. Solving for the steady-state with $\alpha 
> 0$ one immediately arrives at (see \cite{rie05})
\beq
     f(p) \propto p^{-\,(3 + \alpha)}\,\, H(p-p_0)\,,
\eeq where $H(p)$ is the Heaviside step function. Hence for a mean scattering time 
scaling with the gyro-radius (Bohm case), i.e. $\tau \propto p$, $\alpha =1$, one 
obtains $f(p) \propto p^{-4}$ , i.e. a power law particle number density $n(p) 
\propto p^{-2}$ which translates into a synchrotron emissivity $j_{\nu} \propto 
\nu^{-1/2}$.

\section{Possible shear sites in astrophysical jets and associated efficiencies}
In general we may distinguish at least three possible shear sites in astrophysical 
jets, cf. Rieger \& Duffy~(2004,2005):\\
{\bf 1. Longitudinal gradual shear:}
Observationally, there is mounting evidence for an internal velocity stratification 
parallel to the jet axis. In several sources (including 3C353, M87 and Mkn~501, cf. 
\cite{edw00,per99,swa98}), for example, at least a two-component velocity structure 
consisting of a fast velocity spine and a slower moving boundary layer is indicated. 
Such a velocity structure is also indirectly supported by unification arguments for 
BL Lacs and FR~I (\cite{chi00}) and results from hydrodynamical jet simulations 
(e.g., \cite{alo00}). Moreover, Laing et al.~(1999) have argued recently that the 
intensity and polarization systematics in kpc-scale FR~I jets are suggestive of a 
radially (continuously) decreasing velocity profile $v_z(r)$.
In order to estimate the possible particle acceleration efficiency in the presence 
of a longitudinal gradual shear, we may consider a simple realization where the 
flow velocity profile decreases linearly from relativistic to non-relativistic 
speeds over a scale $\Delta r$. It can be shown then (see \cite{rie04}) that the 
minimum acceleration timescale is of the order of
\beq
   t_{\rm acc} \sim \frac{3}{\lambda}\,\frac{(\Delta r)^2}{c\,\gamma_b(r)^4}\,,
\eeq where $\gamma_b(r) >1 $ is the (position-dependent) bulk Lorentz factor of the 
flow and $\lambda$ the particle mean free path. As $\lambda_{\rm proton} \gg
\lambda_{\rm electron}$ the acceleration of electrons is in general much more 
restricted than the acceleration of protons. In the case where the particle mean
free path scales with the gyro-radius, i.e. $\lambda \propto \gamma$ (Bohm case),
the acceleration timescale becomes proportional to $1/\gamma$, and hence reveals 
the same dependency on the particle Lorentz factor $\gamma$ as the synchrotron
(or inverse Compton) cooling timescale. Losses are thus no longer able to stop 
the acceleration process once it has started to work efficiently. 
For illustration, we have plotted in {\bf Fig.~2} the maximum width $\Delta r$ 
allowed for efficient acceleration by balancing $ t_{\rm acc}$ with the 
synchrotron cooling timescale $t_{\rm cool}$, using parameters appropriate for 
the pc-scale radio jets.
 \begin{figure}
    \includegraphics[height=.35\textheight]{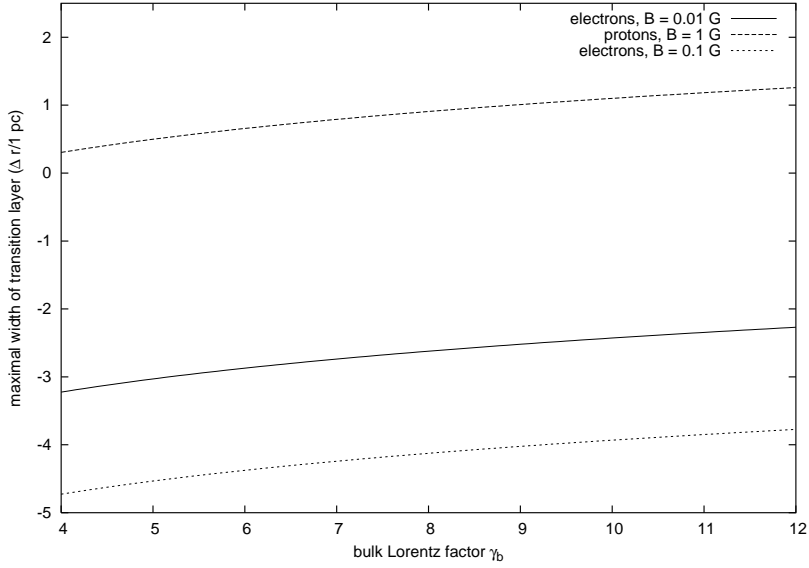}
    \caption{Maximum width of velocity shear layer $\Delta r$ (in parsecs) as a 
     function of the bulk Lorentz factor $\gamma_b$, obtained by requiring the 
     acceleration timescale to be smaller or equal to the synchrotron cooling 
     timescale, i.e. $t_{\rm acc} \leq t_{\rm cool}$. The maximum width $\Delta r$ 
     scales with $B^{-3/2}$ and is proportional to the particle's rest mass square. 
     The acceleration mechanism is generally much more favourable to protons than 
     to electrons.}
      \label{fig2}
  \end{figure}
For a magnetic field strength of $B=0.01$ G, for example, a transition width
$\Delta r_e \leq 0.003$ pc ($\gamma_b = 10$) is required for efficient electron 
acceleration, whereas for protons only $\Delta r_p \leq (m_p/m_e)^2 \,\Delta r_e$ 
is needed.\\
{\bf 2. Longitudinal non-gradual shear:}
When a particle becomes so energetic that its mean free path $\lambda$ becomes 
larger than the width of the velocity transition layer, it will essentially 
experience a non-gradual velocity shear, i.e. a (discontinuous) jump in the
flow velocity. Ostrowski~(1990,1998), for example, has convincingly argued that 
the jet side boundary between the relativistic jet interior and its ambient 
medium may naturally represent a relativistic realization of such a non-gradual 
velocity shear. Based on Monte-Carlo simulations the estimated minimum 
acceleration timescale is of order (cf. \cite{ost90,ost98}).
\beq
   t_{\rm acc} \sim 10\, \frac{r_g}{c} \quad\quad {\rm provided}\quad r_g > 
               \Delta r\,,
\eeq where $r_g$ denotes the gyro-radius of the particle and $\Delta r$ is 
the width of the transition layer. Due to the condition $r_g > \Delta r$
an efficient acceleration of electrons appears excluded by virtue of their
associated rapid radiation losses (e.g., for $B=0.01$ Gauss and $\Delta r 
\geq 0.001$ pc, i.e. $t_{\rm acc} > 10^6$ sec, electron Lorentz factors $\gamma 
\sim 2 \cdot 10^{10}$ would be required in order to fulfill $r_g > \Delta r$, 
implying cooling timescales $t_{\rm cool} \sim 4 \cdot 10^2$ sec well below 
the acceleration timescale). Efficient acceleration of protons, on the other 
hand, can be quite possible as long as their particle mean free paths remain
smaller than the width of the jet.\\
{\bf 3. Transversal gradual shear:}
Apart from a longitudinal velocity stratification, astrophysical jets are also 
likely to have a significant velocity shear perpendicular to their jet axis 
('transversal shear'). In particular, several independent arguments suggest 
that the flow velocity field in these jets is characterized by an additional 
rotational component. The strong correlation between the disk luminosity and 
the bulk kinetic power in the jet (e.g., \cite{raw91}), for example, and the
observational evidence for a disk-origin of jets (e.g., \cite{fal95,mar02}).
suggest that a significant amount of rotational energy of the disk is channeled 
into the jet. Such an internal jet rotation is also implied in theoretical MHD 
models for the origin of jets as magnetized disk winds (e.g., \cite{sau02}). 
In order to evaluate the acceleration potential associated with such shear flow, 
Rieger \& Mannheim~(2002) have recently analyzed the relativistic particle 
transport in rotating and shearing jets. Based on an analytical (mixed-frame) 
approach for the relativistic Boltzmann equation, using a simple (BKG) 
relaxation scattering term, and assuming validity of the diffusion approximation, 
they considered the acceleration of particle in a cylindrical jet model with 
relativistic outflow velocity $v_z$ and different azimuthal rotation profiles. 
In the case of a simple (non-relativistic) Keplerian shear profile they found 
that local power law spectra $f(p) \propto p^{-(3+\alpha)}$ are obtained for
$\tau \propto p^{\alpha}$, $\alpha >0$, whereas for more complex rotation 
profiles (e.g., flat rotation) steeper spectra may be possible.\\ 
In order to gain insights into the acceleration efficiency, let us consider a 
flow velocity field with relativistic $v_z$ and an azimuthal Keplerian rotation 
profile of the form $\Omega(r) = \Omega_k\,(r_{\rm in}/r)^{3/2}$, where $\Omega_k$
is a constant. The acceleration timescale then scales as (see \cite{rie04})
\beq
  t_{\rm acc}(r) \propto \frac{1}{\lambda}\,\frac{1}{\Omega_k^2}
                               \left(\frac{r}{r_{\rm in}}\right)^3\,,
\eeq assuming the flow to be radially confined to $r_{\rm in} \leq r \leq r_j$, 
where $r_j$ is the jet radius, indicating that efficient particle acceleration 
generally requires a region with significant rotation. Note that for a radially 
decreasing rotation profile the higher energy emission will generally be 
concentrated closer to the axis (i.e. toward smaller radii). A comparison of 
acceleration and cooling timescales shows that electron acceleration is usually 
very restricted (i.e., only possible for $r \sim r_{\rm in}$), whereas proton 
acceleration appears well possible. Note again that for $\lambda \propto \gamma$ 
(Bohm case), the acceleration timescales decreases with $\gamma$ in the same way 
as $t_{\rm cool}$, suggesting that losses are no longer able to stop the 
acceleration process once it has started to work efficiently.

\section{Applications}
Observational and theoretical evidence suggest that astrophysical jets may be 
characterized by a significant velocity shear. While internal jet rotation, for 
example, is likely to be present at least in the initial parts of the jet, a 
significant longitudinal velocity shear (parallel to the jet axis) prevailing 
all along the jet might be expected for most powerful sources. The acceleration 
of particles occurring in such shear flows may thus naturally account for a steady 
second population of synchrotron-emitting particles, contributing to the observed 
emission in addition to shock-accelerated ones (cf., \cite{kirk99}). 
Moreover, shear acceleration may also offer a natural explanation for the extended
radio and optical emission observed from several sources. In the case of 3C273, for 
example, the optical spectral index is found to vary only very smoothly along the 
(large-scale) jet in contrast to expectations from simple shock scenarios (cf. 
\cite{jes01}). The jet seems to be highly relativistic even on kpc scales 
(\cite{sam01}) so that longitudinal shear acceleration may perhaps work efficiently 
nearly all along the jet. Moreover, there is also evidence for helical bulk motion 
in the large-scale jet (\cite{bah95}) and internal jet helicity on the VLBI mas-scale 
and below, suggesting that particle acceleration due to internal jet rotation may 
contribute on pc-scale and perhaps even on larger scales.


 \begin{theacknowledgments}
Discussions with John Kirk, Alexandre Marcowith and Lukasz Stawarz, and support by a 
Marie-Curie Individual Fellowship (MCIF - 2002 - 00842) are gratefully acknowledged.
 \end{theacknowledgments}


\bibliographystyle{aipprocl} 

\end{document}